\documentstyle[prl,aps,multicol,epsfig]{revtex}

\def\wid{\end{multicols}
\widetext
\noindent\rule{20.5pc}{0.1mm}\rule{0.1mm}{1.5mm}\hfill}
\def\nar{\hfill\rule[-1.5mm]{0.1mm}{1.5mm}\rule{20.5pc}{0.1mm}
\begin{multicols}{2}
\narrowtext}
\begin{document}
\topmargin -40pt
\draft

\title{Tunneling from 
a correlated 2D electron system transverse to a magnetic field}
\author{T. Sharpee$^{(a)}$, M.I.~Dykman$^{(a)}$, and P.M.~Platzman$^{(b)}$}
\address{$^{(a)}$Department of Physics and Astronomy, 
Michigan State University, East Lansing, Michigan 48824\\ $^{(b)}$Bell
Laboratories, Lucent Technologies, Murray Hill, New Jersey 07974}
\date{\today} 

\maketitle
\begin{quote} 
We show that, in a magnetic field parallel to the 2D electron layer,
strong electron correlations change the rate of tunneling from the
layer exponentially. It results in a specific density dependence of
the escape rate. The mechanism is a dynamical M\"{o}ssbauer-type
recoil, in which the Hall momentum of the tunneling electron is partly
transferred to the whole electron system, depending on the
interrelation between the rate of interelectron momentum exchange and
the tunneling duration. We also show that, in a certain temperature
range, magnetic field can {\it enhance} rather than suppress the
tunneling rate. The effect is due to the magnetic field induced energy
exchange between the in-plane and out-of-plane motion.  Magnetic field
can also induce switching between intra-well states from which the
system tunnels, and a transition from tunneling to thermal activation.
Explicit results are obtained for a Wigner crystal. They are in
qualitative and quantitative agreement with the relevant experimental
data, with no adjustable parameters.
\end{quote} 
\begin{multicols}{2}
\narrowtext 

\section{Introduction}

Many properties of low density two-dimensional electron systems
(2DES) are strongly influenced by electron correlations
\cite{Abrahams-00}-\cite{Spielman-00}. 
Tunneling is one of the most direct tools for revealing these
correlations, as has been demonstrated for systems which display the
quantum Hall effect \cite{DasSarma-97}. In this paper we show that
tunneling can directly reveal correlations in a totally different
class of systems, the low-density 2DES which are far away from the
quantum Hall regime. This happens for tunneling not {\it into} the 2DES, but
{\it from} the 2DES into the volume, and for a magnetic field ${\bf
B}$ applied {\it parallel} rather than perpendicular to the electron
layer.  The effect may not be described in terms of a phenomenological
tunneling Hamiltonian: it is the tunneling matrix element itself that
is sensitive to the electron correlations. As we show, it depends
strongly, and very specifically, on electron density, and also on
temperature and the magnetic field.  An exponentially strong deviation
of the tunneling exponent in a magnetic field from the predictions of
the single-electron theory have been observed for a 2DES on helium
\cite{Andrei-93}. However, these observations remained unexplained.

The field ${\bf B}$ parallel to a 2DES couples the out-of-plane
tunneling motion of an electron to the in-plane motion. For an
isolated electron, which is separated from the continuum by a 1D
potential barrier $U(z)$, see Fig.~1, and is free to move in the plane
$(x,y)$, this results in an exponential suppression of the rate of
tunneling decay.  Indeed, when the electron moves by a distance $z$
away from the layer, it acquires the in-plane Hall velocity $ {\bf
v}_H =(e/c){\bf B}
\times {\bf z}$. The corresponding 
kinetic energy $mv_H^2/2 \equiv m\omega_c^2z^2/2$ is subtracted from
the energy of the out-of-plane tunneling motion ($\omega_c = |eB|/mc$
is the cyclotron frequency), or equivalently, there emerges a
``magnetic barrier'' $m\omega_c^2z^2/2$. This leads to a sharp
decrease of the decay rate.

\begin{figure}
\begin{center}
\epsfxsize=2.2in                
\leavevmode\epsfbox{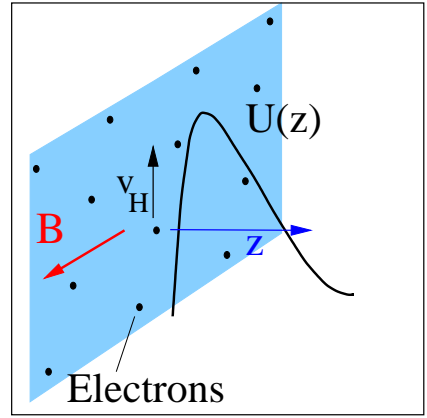}
\end{center}
\caption{The geometry of tunneling from a correlated 2DES transverse 
to a magnetic field; $\omega_p$ is the characteristic in-plane
vibration frequency.}
\label{fig:corr_system}
\end{figure}

The electron-electron interaction can totally change the above
picture. If the electron system is spatially correlated, forming a
Wigner crystal (WC) or a correlated electron liquid, see
Fig.~\ref{fig:corr_system}, the tunneling electron transfers a part of
its in-plane Hall momentum to other electrons. This decreases the loss
of the energy for out-of-plane tunneling motion \cite{Shklovskii}. The
mechanism is similar to that of the M\"ossbauer effect where the
momentum of a gamma quantum is given to the crystal as a
whole. However, in the present case the {\it dynamics} of the
interelectron momentum exchange is very substantial. The
characteristic momentum exchange rate is given by the zone-boundary
plasma frequency $\omega_p$, which is related to the electron density
$n$ by $\omega_p=(2\pi e^2n^{3/2}/m)^{1/2}$. If $\omega_p$ exceeds the
reciprocal duration of underbarrier motion in imaginary time
$\tau_f^{-1}$, the WC momentum adiabatically follows that of the
tunneling electron. As a result, the Hall velocity is the same for all
electrons, and $v_H\propto 1/N \to 0$ ($N$ is the number of
electrons). The effect of the magnetic field on tunneling is then
fully compensated. For $\omega_p\tau_f\sim 1$ the compensation is only
partial. One can say that tunneling is accompanied by creation of
phonons of the WC, and the associated energy goes towards the magnetic
barrier. However, the barrier turns out to be smaller than for a free
electron, and the tunneling rate is then exponentially larger. Still,
for $T=0$ it is much smaller than for $B=0$.

We show in this paper, that unexpectedly, in a certain temperature
range the $B$-induced suppression of the rate of tunneling from a 2DES
may be reversed, and then the decay rate exponentially increases with
$B$. This happens because thermal energy of the in-plane electron
motion is transferred by the magnetic field into the energy of
tunneling motion. Although the effect is generic, as we show below, it
does not arise in systems where the tunneling rate can be found using
the instanton (bounce) technique \cite{Langer}, which is traditionally
applied to describe tunneling for $B=0$\cite{Ao}. There are two reasons which
require to modify this technique. First, the magnetic field
breaks time-reversal symmetry, and therefore, except for the case
where the Hamiltonian of the system has a special form \cite{Leggett},
there are no escape trajectories in real space and imaginary time, and
the system comes out from the barrier with a finite velocity
\cite{us}. Second, for 2D systems the confining potential $U(z)$ is
usually nonparabolic near the minimum, and even nonanalytic, with a
step in the case of heterostructures and, in the case of electrons on
helium, the singularity of the image potential. Therefore such systems
are good candidates for observing magnetic field enhancement of the
tunneling rate.

\begin{figure}
\begin{center}
\epsfxsize=2.5in                
\leavevmode\epsfbox{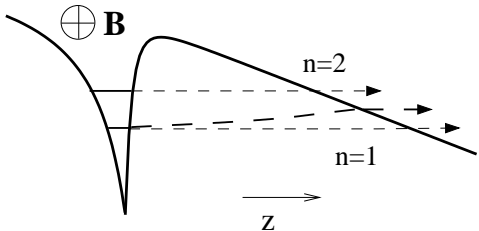}
\end{center}
\caption{Energy transfer from thermal in-plane motion 
into out-of-plane tunneling 
motion induced by the magnetic field (schematically).}
\label{fig:transfer}
\end{figure}

The crossover from suppression to enhancement of tunneling by the
field occurs for a temperature $T_c$ which is of the order of the
reciprocal imaginary duration of underbarrier motion $\tau_0^{-1}$
for $B=0$ in the ground electron state [in what follows we use units
where $\hbar=k_B=1$]. 
If the tunneling rate for $B=0$ is written as $W_0\propto
\exp[-2S_0]$, then $\tau_0= \partial S_0/\partial E_g$, where
$E_g$ is the ground-state energy of out-of-plane motion in the
potential well of $U(z)$.  The tunneling probability increases if the
in plane energy $E_{\rm plane}$ is transferred by the magnetic field
into the tunneling energy $E_{g}$, at least in part, as illustrated in
Fig.~\ref{fig:transfer}. The probability to have an energy $E_{\rm
plane}$ is $\propto
\exp(-E_{\rm plane}/T)$.  Therefore the overall probability, which is
determined by the product of the two probabilities, depends on the
interrelation between $T$ and $\tau_0$.

The time $\tau_0$ also often determines the temperature $T_a$ for
which there occurs a crossover from tunneling decay to decay via
activated overbarrier transitions for $B=0$
\cite{Affleck,Larkin-83}.
Therefore $T_a$ and $T_c$ are of the same order of magnitude. The
interrelation between these temperatures is determined by the
parameters of the system, and various interesting situations may occur
depending on these parameters, as we discuss below. For example, the
logarithm of the escape rate may increase with $B$ even for $T>T_a$,
because in a certain $B$-range, the rate of tunneling from the ground
state exceeds the activation rate, even though it is smaller than the
activation rate for $B=0$. Similarly, with increasing $B$ there may
occur switching from tunneling from the excited intrawell states (see
Fig.~\ref{fig:transfer}) to tunneling from the ground state.

For $T<T_c$, on the other hand, the tunneling rate decreases with the
increasing $B$. However, for large enough $B$ this increase stops,
and escape occurs via thermal activation.

Explicit results on the effect of electron correlations on tunneling
will be obtained assuming that electrons form a Wigner
crystal. Because of strong correlations, overlapping of the wave
functions of individual electrons is small, and electrons can be
``identified''. The problem is then reduced to the tunneling of an
electron coupled to in-plane vibrations of the Wigner crystal. As we
will see, the results provide a good approximation also for a
correlated electron liquid.

In Sec.~II we formulate the model. In Sec.~III we obtain the general
expression for the tunneling rate in the WKB approximation, with
account taken of the discreteness of the energy spectrum of electron
motion transverse to the layer. The result can be understood in terms
of the tunneling trajectory where the duration of motion transverse to
the layer (in imaginary time) is not fixed, it has to be found and
depends on temperature and the magnetic field. In Sec.~IV we analyze
the tunneling exponent, including the case of $T=0$.  In Sec.~V we
discuss temperature effects and show the possibility of $B$-induced
enhancement of tunneling and of switching between different regimes of
escape from the potential well. In Sec.~VI explicit results are
obtained using the Einstein model of the Wigner crystal in which all
phonons are assumed to have the same frequency. Explicit expressions
are obtained for a triangular and square tunneling barriers. In
Sec.~VII we apply the results to electron tunneling from helium
surface and provide a detailed comparison with the experimental data
\cite{Andrei-93}. Sec.~VIII contains concluding remarks.

\section{The model}

A 2D electron system displays strong correlations if the ratio
$\Gamma$ of characteristic Coulomb energy of the electron-electron
interaction $e^2(\pi n)^{1/2}$ to the characteristic kinetic energy is large
(here, $n$ is the electron density). In degenerate systems the kinetic
energy is the Fermi energy $\pi n/m$, whereas in nondegenerate systems
it is the thermal energy $T$. An example of a strongly correlated
nondegenerate 2DES is electrons on helium. The experimental
data for this system refer to the range $\Gamma > 20$
\cite{Andrei_book}.  A classical transition to a Wigner crystal (WC)
was observed for $\Gamma \approx 130$ 
\cite{Wignerexp,Wignerthe}. 
Recently much attention have attracted also strongly correlated
low-density electron and hole systems in semiconductors, where there
have been reached the values of $\Gamma\sim 40$ which are expected to
be sufficient for Wigner crystallization in a degenerate system
\cite{Abrahams-00}.

The effect on tunneling of the magnetic field ${\bf B}$ parallel to
the electron layer is most pronounced if the tunneling length $L$ is
long, because the in-plane Hall momentum due to tunneling $m\omega_c
L$ is simply proportional to $L$. Respectively, of utmost interest to
us are systems with broad and comparatively low barriers. Yet in
experimental systems the barrier widths are most likely to be less
than $ 10^3 \AA$. Therefore, in order to somewhat simplify the analysis
we will assume (although this is not substantial) that $L$ is less
than the average inter-electron distance $\sim n^{-1/2}$. In this
case, since electrons in a strongly correlated system stay away from
each other, the in-plane electron dynamics only weakly affects the
tunneling potential \cite{AP-90}. We will neglect the corresponding
coupling for $B=0$. The major effect on
tunneling comes from a few nearest neighbors, and the presence or
absence of long-range order in the 2DES does not affect the tunneling
rate. Therefore we will analyze tunneling assuming that the electron
system is a Wigner crystal. As we will see, the results will indeed
depend on the short-wavelength modes of the WC, as expected from the
above arguments, and therefore we believe that the model provides a
good approximation even where electrons form a correlated fluid.

In a strongly correlated system, exchange effects are not significant,
and one can identify the tunneling electron. Its out-of-plane motion
for $B=0$ is described by the Hamiltonian
\begin{equation}
\label{H0}
H_0={p_z^2 \over {2 m}}+U(z).
\end{equation}
The potential $U(z)$ has a well which is separated by a tunneling
barrier from the extended states with a quasi-continuous spectrum,
cf. Fig.~\ref{fig:corr_system}. The well is nonparabolic near the
minimum, in the general case. The metastable intrawell states are
quantized. We will consider temperatures for which nearly all
electrons are in the lowest level, with energy $E_g$.

The magnetic field ${\bf B}$ parallel to the layer mixes the
out-of-plane motion of the tunneling electron with the in-plane
vibrations of the Wigner crystal. The full Hamiltonian is of the form
\begin{equation}
\label{hamiltonian}
H=H_0 +H_B +H_v,
\end{equation} 
with
\begin{equation}
\label{Hv}
H_v= {1\over 2}\sum_{{\bf k},j}\left[m^{-1}{\bf p}_{{\bf k}j}{\bf
p}_{-{\bf k}j} + m \omega_{{\bf k}j}^2 {\bf u}_{{\bf k}j}{\bf
u}_{-{\bf k}j}\right]
\end{equation}
and
\begin{equation}
\label{interaction}
H_B={1 \over 2} m\omega_c^2 z^2 - \omega_c z N^{-1/2}
\sum_{{\bf k},j}[\hat{\bf B}\times{\bf p}_{{\bf k}j}]_z.
\end{equation}
Here, ${\bf p}_{{\bf k}j}$, ${\bf u}_{{\bf k}j}$, and $\omega_{{\bf
k}j}$ are the 2D momentum, displacement, and frequency of the WC
phonon of branch $j$ ($j=1,2$) with a 2D wave vector ${\bf k}$.  We
chose the equilibrium in-plane position of the tunneling
electron to be at the origin. Then its in-plane 2D momentum is ${\bf p}
=N^{-1/2}\sum{\bf p}_{{\bf k}j} $ for $B=0$.

The interaction Hamiltonian $H_B$ (\ref{interaction}) {\it does not}
conserve the phonon quasi-momentum ${\bf k}$. The Hall momentum of the
tunneling electron is transferred to the WC as a whole.  The term
$H_B$ couples the out-of-plane motion to lattice vibrations. The
problem of many-electron tunneling is thus mapped onto a familiar
problem of a particle coupled to a bath of harmonic oscillators
\cite{Feynman,Leggett}, with the coupling strength controlled by the
magnetic field. The distinctions from the standard situation stem from
the non-parabolicity of the potential well near the minimum and from
the fact that coupled by $H_B$ are the electron {\it coordinate} $z$
and the in-plane {\it momenta} of the lattice.  These quantities have
different symmetry with respect to time inversion. In the general case
[for example, where the potential energy of the system has odd-order
terms in the displacements ${\bf u}_{{\bf k}j}$], the broken
time-reversal symmetry requires a special approach to the analysis of
tunneling \cite{us}. The results discussed below can be appropriately
generalized using this approach.

For the model (\ref{hamiltonian}), the analysis is simplified by the
structure of the Hamiltonian (cf. \cite{Leggett}). For vibrations with
the Hamiltonian $H_v$ (\ref{Hv}), one can always make a canonical
transformation from the canonical coordinates and momenta ${\bf
u}_{{\bf k}j}$ and ${\bf p}_{{\bf k}j}$ to the new canonical
coordinates and momenta ${\bf p}_{{\bf k}j}$ and $
-{\bf u}_{{\bf k}j}$, respectively. This transformation
interchanges the time-reversal symmetry of the in-plane dynamical
variables, it makes ${\bf p}_{{\bf k}j}$ and ${\bf u}_{{\bf k}j}$ even
and odd in time, respectively. Because $H_B$ is independent of ${\bf
u}_{{\bf k}j}$ and is linear in ${\bf p}_{{\bf k}j}$, in the new
variables it takes on a more familiar form of a ``potential''
coupling, with energy which depends on dynamical coordinates only and
with ``restored'' time-reversal symmetry.

\section{The WKB approximation}
\subsection{General formulation}
We will evaluate the escape rate $W$ in the WKB approximation. The
major emphasis will be placed on the tunneling exponent. We will
assume that the escape rate is much less than the intrawell relaxation
rate for relevant states, and there is formed thermal
distribution over the intrawell states of the system. This is not
necessarily true for 2D systems. Our results can be generalized to the
case of slow intrawell relaxation, see Sec.~VII.

Electron-phonon interaction under the barrier is strong. One should
therefore think of escape of the coupled electron-phonon system. It
results from decay of the metastable intrawell states $\alpha$, with
decay rates $W_{\alpha}$.
These rates sharply increase with state energies $E_{\alpha}$, whereas
the Boltzmann intrawell distribution exponentially decreases with
$E_{\alpha}$. As a result, there is a comparatively small group of
states which mostly contribute to the escape (for fast intrawell
relaxation, the relataive population of these states remains
unchanged). This allows one to characterize escape by a single
rate $W$,

\begin{eqnarray}
\label{W_as_a_sum}
&&W =Z^{-1}\sum\nolimits_{\alpha}W_{\alpha}\exp(-\beta E_{\alpha}),\\
&&W_{\alpha}= C_{\alpha}
\exp\left[-2S_{\alpha}(\bbox{\xi}_{\rm f},\bbox{\xi}_{\rm
in})\right]
\left|\psi_{\alpha}(\bbox{\xi}_{\rm in})\right|^2.\nonumber
\end{eqnarray}
Here, we introduced a vector $\bbox{\xi}=\left(z,\{ {\bf p}_{{\bf k}j}
\} \right)$ with components which enumerate the
$z$-coordinate of the tunneling electron and the ``coordinates'' ${\bf
p}_{{\bf k}j}$ of the phonons, $Z$ is the partition function
calculated neglecting escape, and $C_{\alpha}$ are the prefactors in
the partial escape rates, they will not be discussed in this paper.

The exponents in the rates $W_{\alpha}$ are determined
\cite{Landau-QM} by the wave functions
$\psi_{\alpha}(\bbox{\xi})$ at the turning points $\bbox{\xi}_{\rm f}$
on the boundary of the classically accessible ranges ($\bbox{\xi}_{\rm
f}$ depend on $\alpha$, see below). It is convenient to evaluate
$\psi_{\alpha}(\bbox{\xi}_{\rm f})$ in two steps, each of which gives
an exponential factor.  The first factor,
$\exp[-S_{\alpha}(\bbox{\xi}_{\rm f},\bbox{\xi}_{\rm in})]$, describes
the decay of the wave function under the barrier. Formally, it relates
$\psi_{\alpha}(\bbox{\xi}_{\rm f})$ to $\psi_{\alpha}(\bbox{\xi}_{\rm
in})$. The point $\bbox{\xi}_{\rm in}$ is chosen close to the well,
yet it lies under the barrier, so that $S_{\alpha}$ can be calculated
in the WKB approximation. The second factor is
$\psi_{\alpha}(\bbox{\xi}_{\rm in})$ itself. The resulting rate should
be independent of $\bbox{\xi}_{\rm in}$.

We start with the function $S_{\alpha}(\bbox{\xi},\bbox{\xi}_{\rm
in})$. To the lowest order in $\hbar$, for systems with time-reversal
symmetry (which we ``restored'' by the canonical transformation) it is
the action for a classical underbarrier motion in imaginary time $\tau
= it$ with purely imaginary momenta \cite{Langer}

\begin{equation}
\label{momenta}
p_z = i\,\partial S/\partial z, \quad
{\bf u}_{{\bf k}j} = -i\,\partial S/\partial {\bf p}_{{\bf k}j}.
\end{equation}
As a function of the imaginary time $\tau$, the action
$S(\bbox{\xi},\bbox{\xi}_{\rm in})$ is given by the integral of the Euclidean
Lagrangian $L_{\rm E}$,
\begin{equation}
\label{S_E}
S_{\alpha}(\bbox{\xi},\bbox{\xi}_{\rm in})= \int_0^{\tau} L_{\rm E} d\tau
-E_{\alpha}\tau,
\end{equation}
The Lagrangian $L_{\rm E}$ is obtained from the Hamiltonian
(\ref{hamiltonian}) using the Legendre transformation $L=p_z(dz/dt) -
\sum {\bf u}_{{\bf k}j}(d{{\bf p}}_{{\bf k}j}/dt)-H$], followed 
by the transition to imaginary time, which gives
\begin{equation}
\label{Lagrangian}
L_{\rm E}=L_0+L_v+L_B.
\end{equation}
Here,
\begin{equation}
L_0={m\over 2} \left({dz \over d\tau}\right)^2+U(z), \quad L_B=H_B,
\label{L_0}
\end{equation}
and $L_v$ is the Lagrangian of the phonons, $L_v=\sum\nolimits_{{\bf
k}j}L_{{\bf k}j}$, with
\begin{equation}
\label{L_v}
L_{{\bf k}j}= {1\over 2m} {\bf
p}_{{\bf k}j}{\bf p}_{{\bf -k}j} + {1\over 2m\omega_{{\bf k}j}^2} {d
{\bf p}_{{\bf k}j} \over d\tau} {d {\bf p}_{{\bf -k}j} \over d\tau}.
\end{equation}

The classical equations of motion in imaginary time have the standard form

\begin{equation}
\label{eom}
{d\over d\tau}{\partial L_{\rm E}\over \partial \dot{\bbox{\xi}}}  - 
{\partial L_{\rm E}\over \partial \bbox{\xi}}=0.
\end{equation}
where overdot means differentiation over $\tau$.  To calculate the
escape rate, one has to find the trajectory which goes from
$\bbox{\xi}(0)=\bbox{\xi}_{\rm in}$ to the boundary of the classically
accessible range $\bbox{\xi}_{\rm f}$ at a certain time $\tau_f$ and calculate
the action $S_{\alpha}$ along this trajectory.

If the potential barrier $U(z)$ is smooth, the wave function and its
derivatives under the barrier have to match the WKB wave function in
the classically allowed range behind the barrier. The matching occurs
at a turning point of the classical motion (\ref{eom}) where the
derivatives of the both wave functions become equal to zero
\cite{Landau-QM}, i.e. for $\partial S_{\alpha}/\partial z = \partial
S_{\alpha}/\partial {\bf p}_{{\bf k}j} = 0$, i.e.
\begin{equation}
\label{bound_Euclid}
\dot z(\tau_f)=0,\; \dot {\bf p}_{{\bf k}j}(\tau_f)={\bf 0}.
\end{equation}
Eq.~(\ref{bound_Euclid}) is also the condition of the extremum of
$S_{\alpha}$ with respect to the points $\bbox{\xi}$ on the boundary
of the classically accessible range: the escape rate is determined by
the minimum of $S_{\alpha}$ on this boundary. A detailed analysis of
the behavior of multidimensional tunneling trajectories in imaginary
time for systems with time-reversal symmetry is given in
Ref.~\cite{Schmid}.

Time-reversal symmetry of the equations (\ref{eom}) in coordinates
$(z, {\bf p}_{{\bf k}j})$, together with the condition
(\ref{bound_Euclid}), show that, if the equations of motion are
extended beyond $\tau_f$, the system will bounce off the turning point
and then move under the barrier back to the starting point. The
section of the trajectory for $\tau >
\tau_f$ is mirror-symmetrical to that for  $\tau < \tau_f$,
\begin{equation}
\label{symmetry}
z(\tau_f+ \tau)= z(\tau_f-\tau), \, {\bf p}_{{\bf k}j}(\tau_f+ \tau)= {\bf
p}_{{\bf k}j}(\tau_f-\tau),
\end{equation}
where $ 0\leq \tau \leq \tau_f$. As a result, the tunneling exponent
$2S_{\alpha}$ can be calculated along the trajectory
(\ref{eom}) that reaches the turning point at $\tau_f$ and returns to
the well at $2\tau_f$.

The time $\tau_f$ is determined by the boundary conditions
(\ref{bound_Euclid}) and by the initial conditions on the trajectory,
which are given by $\psi_{\alpha}(\bbox{\xi}_{\rm in})$. 
If the intrawell dynamics is semiclassical, the dominating
contribution to the overall rate $W$ (\ref{W_as_a_sum}) comes from the
energies $E_{\alpha}$ for which the duration of the tunneling motion
$\tau_f=\beta/2$
\cite{Langer}. In the general case this is no longer true.

\subsection{The wave function close to the well}

We are interested in the case where the width of the quantum well is
much less than the typical width $L$ of the tunneling barrier.  More
generally, we assume that for low-lying intrawell states $n$, the
characteristic lengths $1/\gamma_n$ of localization in the
$z$-direction are $\gamma_n^{-1}\ll L$.  Then, even where the effect
of the magnetic field accumulates under the barrier and the tunneling
rate is strongly changed,
the field may still only weakly perturb the intrawell motion. In this
case, inside the well and close to it, the  out-of-plane electron
motion is separated from the in-plane vibrations. Respectively, the states
of the electron-phonon system can be enumerated by $n$ and the phonon
occupation numbers $n_{{\bf k}j}$, i.e. $\alpha = (n, \{n_{{\bf
k}j}\})$, and the energies are
\begin{equation}
\label{energies}
E_{\alpha}= E_n + \sum\nolimits_{{\bf k}j}\varepsilon_{{\bf k}j},\quad
\varepsilon_{{\bf k}j}=\omega_{{\bf k}j}n_{{\bf k}j}.
\end{equation}
Usually the interlevel distances $E_{n+1}-E_n \gg \omega_{{\bf k}j}$,
for low-lying levels.

Because of the separation of motions, we can choose a plane $z=z_{\rm
in}$ under the barrier but close to the well, so that for $\bbox{\xi}
\approx \bbox{\xi}_{\rm in}$ the wave functions
$\psi_{\alpha}(\bbox{\xi})$ are semiclassical and at the same
time can be factored,

\begin{equation}
\label{factored_psi}
\psi_{n,\{n_{{\bf k}j}\}}(\bbox{\xi}) \propto e^{-\gamma_n z}
\exp\left[-\sum\nolimits_{{\bf k}j}S_{n_{{\bf k}j}}({\bf p}_{{\bf k}j})\right].
\end{equation}
The action $S_{n_{{\bf k}j}}$ determines the dependence of the wave
function on the phonon coordinates.

For $\bbox{\xi}=\bbox{\xi}_{\rm in}$, Eq.~(\ref{factored_psi}) gives
the initial values of the dynamical variables $\bbox{\xi}(0)\equiv
\bbox{\xi}_{\rm in}$ and $\dot{\bbox{\xi}}(0)$ on the WKB trajectory
(\ref{eom}). In particular if, for $z\approx z_{\rm in}$, the
potential $U(z)$ varies over the distance much bigger than
$1/\gamma_n$, then
\begin{equation}
\label{initial1}
z(0) = z_{\rm in},\quad \dot z(0) = {\gamma_n\over m} =
\left[{2[U(z_{\rm in})-E_n]\over m}\right]^{1/2},
\end{equation} 
and $\gamma_n$ (\ref{initial1}) is independent of the exact position of
the plane $z=z_{\rm in}$.

It is convenient to write $S_{n_{{\bf k}j}}$ and ${\bf p}_{{\bf k}j}$
in Eq.~(\ref{factored_psi}) in the energy-phase representation, using
the phonon energy $\varepsilon_{{\bf k}j}$ and the imaginary time
$\tau_{{\bf k}j}$ it takes for a phonon to move under the barrier from
the boundary $(2m\varepsilon_{{\bf k}j})^{1/2}$ of the classically
allowed region to the given ${\bf p}_{{\bf k}j}$.  With the Euclidean
Lagrangian of the phonons (\ref{L_v}), we have for ${\bf p}_{{\bf
k}j}=[{\bf p}_{{\bf k}j}]_{\rm in}\equiv {\bf p}_{{\bf k}j}(0)$
\begin{equation}
\label{initialS}
S_{n_{{\bf k}j}}\left({\bf p}_{{\bf k}j}(0)\right)
= \int_{-\tau_{{\bf k}j}}^0 d \tau L_{{\bf k}j}
(\tau)-\varepsilon_{{\bf k}j}\tau_{{\bf k}j},
\end{equation}
and
\begin{eqnarray}
\label{initial_p_u}
&&{\bf p}_{{\bf k}j}(0)={\bf e}_{{\bf k}j}(2m\varepsilon_{{\bf
k}j})^{1/2}\cosh\omega_{{\bf k}j}\tau_{{\bf k}j},\\ 
&&\dot{\bf
p}_{{\bf k}j}(0)= {\bf e}_{{-\bf k}j}\left(2\varepsilon_{{\bf
k}j} m\omega^2_{{\bf k}j}\right)^{1/2}
\sinh\omega_{{\bf k}j}\tau_{{\bf k}j}
\nonumber
\end{eqnarray}
[${\bf e}_{{\bf k}j}$ is the polarization vector of the mode $({\bf
k},j)$].

\subsection{A three-segment optimal trajectory}

To evaluate the escape rate $W$ to logarithmic accuracy, one can,
following Feynman's procedure, solve the equations of motion
(\ref{eom}) for the vibration ``coordinates'' ${\bf p}_{{\bf
k}j}(\tau)$ in terms of $z(\tau)$ and the initial energies
$\varepsilon_{{\bf k}j}$ and phases $\omega_{{\bf k}j}\tau_{{\bf
k}j}$. Then, from the boundary condition (\ref{bound_Euclid}), one can
express $\tau_{{\bf k}j}$ in terms of other variables, and then perform
thermal averaging by summing the escape rate over
$n_{{\bf k}j}$ with the Boltzmann weighting factor. Here we
give an alternative derivation, which provides a better insight into
the structure of the tunneling trajectory.

We note that, from Eqs.(\ref{W_as_a_sum}), (\ref{S_E}),
(\ref{factored_psi}), and (\ref{initialS}), the partial escape rate
$W_{\alpha}$ can be written as $W_{\alpha}\propto \exp(-s_{\alpha})$, with
\wid
\begin{eqnarray}
\label{W_alpha}
 s_{\alpha} = &&
\sum\nolimits_{{\bf k}j}
\int\nolimits_{-\tau_{{\bf k}j}}^0 d \tau\, L_{{\bf k}j}
(\tau)
 + \int_0^{2\tau_f}d\tau \,L_{\rm E}(\tau) + \sum\nolimits_{{\bf k}j}
\int\nolimits_{2\tau_f}^{2\tau_f+\tau_{{\bf k}j}} d \tau\, L_{{\bf k}j}
(\tau)
-2E_n\tau_f - 2\sum\nolimits_{{\bf
k}j}\varepsilon_{{\bf k}j}(\tau_f+\tau_{{\bf k}j}).
\end{eqnarray}
\nar
\noindent
(the term $\gamma_n z_{\rm in}$ in (\ref{factored_psi}), which is
small compared to $s_{\alpha}\sim \gamma_n L$, is incorporated into the
prefactor, see Sec.~VII~A).

Eq.~(\ref{W_alpha}) can be interpreted in the following way: the
tunneling electron in its $n$th state, accompanied by phonons, move
under the barrier along a classical trajectory for the imaginary time
$2\tau_f$. This motion is described by the Lagrangian $L_{\rm
E}$. Before and after that, the phonons are moving on their own,
disconnected from the electronic $z$-motion, for times $\tau_{{\bf
k}j}$ and with the Lagrangian $L_v$, so that the overall phonon
trajectories make closed loops which start and end at turning points.

\begin{figure}
\begin{center}
\epsfxsize=3.35in                
\leavevmode\epsfbox{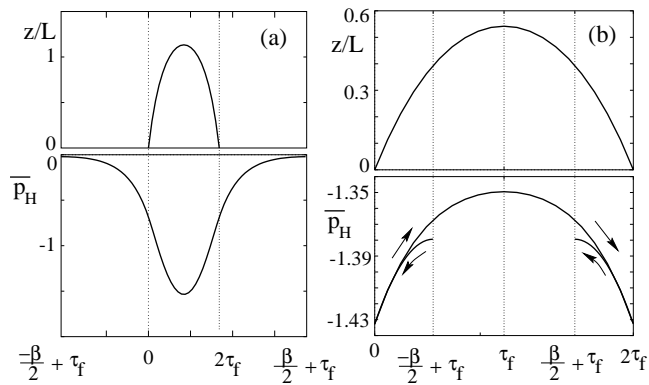}
\end{center}
\caption{Optimal  trajectories of the tunneling electron
$z(\tau)$ and the vibrations of the WC for $\beta> 2\tau_f$ (a) and
$\beta<2\tau_f$ (b). The numerical data refer to the Einstein model of
the Wigner crystal, with $\overline{p_H}$ being the vibrational momentum in
the Hall direction $\hat {\bf z}\times {\bf B}$, in units
$\hbar\gamma/2$. The arrows show the direction of motion along the
optimal trajectory when $\beta<2\tau_f$. The tunneling potential is of
the form (\protect\ref{triangular}), with dimensionless cyclotron
frequency $\omega_c\tau_0=2.0$, where $\tau_0=2mL/\gamma$ is the
imaginary transit time for $B=0$.  The vibrational frequency is
$\omega_p\tau_0=1.0$.}
\label{fig:transition.eps}
\end{figure}

In the WKB approximation, the sum over the phonon occupation numbers
$n_{{\bf k}j}$ of the weighted partial probabilities $W_{\alpha}$
(\ref{W_as_a_sum}) can be replaced by the integral over
$\varepsilon_{{\bf k}j}$, which should then be evaluated by the
steepest descent. The corrections due to the discreteness of the
values of $\varepsilon_{{\bf k}j}$ are small provided $\omega_{{\bf
k}j}\tau_f \ll s_{\alpha}$. From (\ref{W_alpha}), the extremum of
$\exp(-s_{\alpha}-\beta E_{\alpha})$ with respect to
$\varepsilon_{{\bf k}j}$ is reached for

\begin{equation}
\label{tau}
\tau_{{\bf k}j} ={1\over 2} \beta - \tau_f,
\end{equation}
This expression shows that the duration of the free phonon motion
$\tau_{{\bf k}j}$ is the same for all vibrational modes. Moreover, the
overall duration of the three-segment optimal trajectory of each
vibration is $2(\tau_{{\bf k}j}+\tau_f)=\beta$. Examples of the
trajectories are shown in Fig.~\ref{fig:transition.eps}.

For low temperatures, $\beta > 2\tau_f$, the direction of time along
the vibrational trajectory does not change, $\tau_{{\bf k}j} > 0$. The
corresponding branch of the intrawell vibrational wave function
(\ref{factored_psi}) $\propto\exp(-S_{n_{{\bf k}j}}[{\bf p}_{{\bf
k}j}(0)])$ decays with the increasing $|{\bf p}_{{\bf k}j}(0)|$ in the
classically forbidden region $|{\bf p}_{{\bf k}j}(0)|>
(2m\varepsilon_{{\bf k}j})^{1/2}$. In contrast, for $\beta < 2\tau_f$,
we have $\tau_{{\bf k}j}<0$. This shows that the extremum over
$\varepsilon_{{\bf k}j}$ is reached if the intrawell vibrational wave
function is analytically continued from the decaying to the increasing
branch.

For $\tau_{{\bf k}j} = (\beta/2)-\tau_f < 0$, the ``free-vibrations''
term $S_{n_{{\bf k}j}}$ is negative, it gives rise to the decrease of
the tunneling exponent. This is the formal reason why, for
sufficiently high temperatures, an in-plane magnetic field can
increase the tunneling rate compared to its $B=0$ value by coupling
thermally-excited in-plane vibrations to tunneling.

If the intrawell motion transverse to the layer were semiclassical,
the sum over the energy levels of this motion $E_n$ in
Eq.~(\ref{W_as_a_sum}) could be replaced by an integral. The extremum
of the integrand is reached for
$\tau_f=\beta/2, \tau_{{\bf k}j}=0$. This is the familiar result of
the instanton theory, in which the whole system moves under the
barrier from the well to the turning point and back over the imaginary
time $\beta$ \cite{Langer}. Clearly, in this case one should not
expect the tunneling rate to be enhanced by a magnetic field.

In the case of 2D electron systems, the potential well is {\it not}
parabolic, and each term in the sum over $n$ (\ref{W_as_a_sum}) has to
be considered separately. Except for narrow parameter intervals, the
contribution of one of them is dominating, and the electron tunnels
from the corresponding state.

\section{The tunneling exponent}

Eqs.~(\ref{W_as_a_sum}) and (\ref{W_alpha}) allow us to write the
escape rate as a sum of the escape rates for different intrawell
states $n$. To logarithmic accuracy
\begin{equation}
\label{general_W}
W\propto \max_n\,\exp(-R_n), \quad R_n = \min_{z(\tau)}
{\cal R}_n[z(\tau)],
\end{equation}
The functional ${\cal R}_n[z]$ is a retarded action functional for a 1D
motion normal to the electron layer. It is determined by the
functional $s_{\alpha}$ for the $n$th state from which the
the dynamical variables of the in-plane vibrations have been
eliminated in a standard way
\cite{Feynman} by solving the linear equations of motion (\ref{eom})
for ${\bf p}_{{\bf k}j}(\tau)$, with account taken of the expression
for the duration of the phonon tunneling motion (\ref{tau}). This
gives
\begin{eqnarray}
\label{tun_Exp}
&{\cal R}_n[z]=&\int_0^{2\tau_f}d\tau_1 \left[ {m\over 2}\left({dz\over
d\tau}\right)^2+U(z) +{1\over 2}m\omega_c^2 z^2
(\tau_1)\right]\nonumber \\ 
&& + {\cal R}_{\rm ee}[z] +(\beta-2\tau_f)E_n.
\end{eqnarray}
(we count $z$ off from $z_{\rm in}$, i.e. $z_{\rm in}=0$).

The term ${\cal R}_{\rm ee}$ gives the retarded action which results from the
electron-electron interaction,

\begin{eqnarray}
\label{retarded}
{\cal R}_{\rm ee}[z]= -&&{m \omega_c^2 \over 4N}\sum_{{\bf
k}j}
\omega_{{\bf k}j}\int_0^{2\tau_f}\int_0^{\tau_1} d\tau_1 d
\tau_2 \nonumber\\
&&\times z(\tau_1)z(\tau_2) \chi_{{\bf k}j}(\tau_1-\tau_2)
\end{eqnarray}
The
function $\chi_{{\bf k}j}(\tau)$ is the phonon Green's function,
\[
\chi_{{\bf k}j}(\tau)= \bar{n}_{{\bf k}j} \exp [\omega_{{\bf k}j}\tau]+
(\bar{n}_{{\bf k}j}+1)\exp
[-\omega_{{\bf k}j} \tau]
\]
($\bar{n}_{{\bf k}j}= [ \exp ( \beta \omega_{{\bf k}j}) -1]^{-1}$ is
the thermal occupation number). We note that ${\cal R}_{\rm ee}[z]$
can be also expressed in terms of the correlation function
$\tilde\chi(\tau)=\langle {\bf p}_{\parallel}(\tau){\bf
p}_{\parallel}(0)\rangle$ of the in-plane momentum ${\bf
p}_{\parallel}$ of an electron in the correlated 2DES,
\begin{eqnarray}
\label{alternative_retarded}
{\cal R}_{\rm ee}[z] =  -{1\over 2}\omega_c^2
\int_0^{2\tau_f}\! \int_0^{\tau_1} d\tau_1 d
\tau_2 \, z(\tau_1)z(\tau_2)\tilde\chi(\tau_1-\tau_2).
\end{eqnarray}
We expect that not only does this expression apply to a Wigner
crystal, but it also provides a good approximation in the case of a
correlated electron liquid; it corresponds to the lowest-order term in
the cumulant expansion of the appropriate propagator.

The term ${\cal R}_{\rm ee}$ is negative. It means that the electron-electron
interaction in a correlated 2DES always {\it increases} the tunneling
rate in the presence of a magnetic field.  Moreover, when this term
exceeds $(m\omega_c^2/2)\int z^2 d\tau$, the tunneling exponent as a
whole decreases with the increasing $B$. 

Two physical phenomena are described by the term ${\cal R}_{\rm ee}$. One is
the dynamical compensation of the Hall momentum of the tunneling
electron by the WC as the electron moves under the barrier in the
$z$-direction. The other is thermal ``preparation'' of the Hall
momentum for the tunneling electron, which is then transformed by the
magnetic field into the momentum of motion in the $z$-direction. We
analyze these effects in the following subsections.

\subsection{Zero temperature limit}

It would be natural to think that, since tunneling is accompanied by
creation of phonons for $T=0$, then the higher the phonon frequency
the lower the tunneling rate. In fact just the opposite is true.

The effect of the electron-electron interaction on tunneling, as
characterized by ${\cal R}_{\rm ee}$, depends on the interrelation
between the characteristic phonon frequency $\omega_p$ and the
tunneling duration $\tau_f$.  When the tunneling electron is
``pushed'' by the Lorentz force, it exchanges the in-plane momentum
with other electrons. The parameter $\omega_p\tau_f$ determines what
portion of the momentum goes to the crystal as a whole during the
tunneling (note that the tunneling motion goes in imaginary time, and
the quantity $\tau_f$ characterizes the time uncertainty rather than
the actual duration of a real process, see
Ref.~\cite{Landauer-Martin}). As mentioned in the Introduction, in the
adiabatic limit of large $\omega_p\tau_f$, all electrons have same
in-plane velocity, with an accuracy to quantum fluctuations. Therefore
the Lorentz force produces no acceleration, and {\it no phonons} are
created during the tunneling. The effect of the magnetic field on
tunneling should then be eliminated.

These arguments are confirmed by the analysis of
Eq.~(\ref{retarded}). If the electron system is rigid enough in the
plane, so that $\omega_{{\bf k}j}\tau_f\gg 1$, the major contribution
to ${\cal R}_{\rm ee}$ comes from $\tau_1-\tau_2 \sim \omega_{{\bf
k}j}^{-1} \ll \tau_f$. Therefore $z(\tau_2)\approx z(\tau_1)$, so that
in ${\cal R}_{\rm ee}$ the two terms $\propto \omega_c^2$ compensate
each other. The tunneling occurs as if the electron is disconnected
from the phonons and does not experience a magnetic field. However, a
simple analysis shows that the electron mass is effectively
incremented by a $B$-dependent factor, and the tunneling exponent
$R\equiv R_1$ is appropriately renormalized,
\begin{eqnarray}
\label{m*}
m\to m^*,\;&& m^*=m\left[1+(2N)^{-1}\sum\nolimits_{{\bf
k}j}\left(\omega_c^2/ \omega^2_{{\bf k}j}\right)\right], \\
&& R = \left( m^*/m\right)^{1/2}R_{B=0}.\nonumber
\end{eqnarray}
Here, the sum is limited from below by the condition $\omega_{{\bf
k}j} > \tau_f^{-1}$; for a Wigner crystal, the dependence of the mass
renormalization on the cutoff frequency is logarithmic. The tunneling
rate approaches its value for $B=0$ with increasing $\omega_p$. On the
other hand, the slope of the logarithm of the tunneling rate as a
function of $\omega_c$ depends explicitly on $\omega_p$, for $\omega_c
\agt\omega_p$. This provides a means for measuring $\omega_p$.

For $\omega_p\tau_f\sim 1$, only a part of the Hall momentum can be
taken by the electron crystal. The rest goes into the in-plane kinetic
energy of the tunneling electron, and ultimately into creations of WC
phonons. However, the major contribution to ${\cal R}_{\rm ee}$ still comes
from high-frequency phonons. It can be shown from (\ref{retarded})
that this contribution monotonically increases with increasing
$\omega_{{\bf k}j}$. This is because the more rigid the electron
system is, the more effectively it compensates the in-plane Hall
momentum. An important consequence is that, since high-frequency
vibrations have small wavelengths, the major effect on tunneling comes
from short-range order in the electron system.

On the whole, for $T=0$, the magnetic-field induced term in the
tunneling exponent is positive, i.e. the tunneling rate decreases with
the magnetic field. This can be seen from Eqs.~(\ref{tun_Exp}),
(\ref{retarded}) by replacing $z(\tau_1)z(\tau_2)$ in ${\cal R}_{\rm ee}$
with $(1/2)[z^2(\tau_1) + z^2(\tau_2]\geq z(\tau_1)z(\tau_2)$ and then
integrating the function $\chi(\tau_1-\tau_2)=\exp[-\omega_{{\bf
k}j}(\tau_1-\tau_2)]$ over $\tau_2$ [for the term $z^2(\tau_1)$] or
$\tau_1$ [for $z^2(\tau_2)$].  

Electron correlations exponentially reduce the effect of the magnetic
field on the tunneling rate in a magnetic field.  For specific models,
the dependence of the tunneling rate on $B$ and the vibration
frequencies will be illustrated in Sec.~VI, and the results will be
compared with the experiment.

\subsection{Small phonon frequencies}

The analysis of the tunneling rate somewhat simplifies in the case of
comparatively high temperatures and small phonon frequencies, where
the vibrations are classical and their frequencies are small compared
to the reciprocal tunneling duration, $\omega_{{\bf k}j}\beta,
\omega_{{\bf k}j}\tau_f \ll 1$. In this case

\begin{equation}
\label{classical}
{\cal R}_{\rm ee}[z] = -2mT\omega_c^2\tau_f^2\bar z^2,\quad \bar
z =\tau_f^{-1}\int_0^{\tau_f} d\tau z(\tau)
\end{equation}
[since we chose $z_{\rm in}=0$ and $z_f >0$, we have ${\cal
R}_{\rm ee} < 0$ in Eq.~(\ref{classical})].

Eqs.~(\ref{tun_Exp}), (\ref{classical}) describe also the
tunneling action of a single electron, with the Maxwell distribution
of the in-plane momentum inside the well $\exp(- p^2/2mT)$. The
coupling of the $\hat{\bf z}\times {\bf B}$ component of the momentum
to the out-of-plane motion gives rise to the term
$-2p\omega_c\int_0^{\tau_f}d\tau z(\tau)$ in the tunneling action
[cf. Eqs.~(\ref{interaction}), (\ref{S_E})]. The extreme value of the
sum of this term and $-p^2/2mT$ is just equal to $-{\cal R}_{\rm ee}[z]$ as
given by (\ref{classical}). 

The single-electron form of the tunneling exponent is to be
expected in the limit of small $\omega_{{\bf k}j}$, because the
distribution over in-plane momenta of electrons forming a Wigner
crystal is Maxwellian, in the classical limit. For small $\omega_{{\bf
k}j}\tau_f$ the momenta do not change over the tunneling duration,
therefore only the momentum of the tunneling electron itself is
important.  The above derivation provides an independent test of the
derivation used to obtain the general expression (\ref{tun_Exp}),
(\ref{retarded}).

We note that the action ${\cal R}_{\rm ee}[z]$ (\ref{classical}) is still
retarded, it does not correspond to a local in time Lagrangian. The
functional form of ${\cal R}_{\rm ee}$ remains the same even for temperatures
$T\alt \omega_{{\bf k}j}$ provided the phonon frequencies are small
compared to $\tau_f^{-1}$ and $\omega_c$. In this case
$T$ in Eq.~(\ref{classical}) has to be replaced by
$(4N)^{-1}\sum\omega_{{\bf k}j}(2\bar n_{{\bf k}j} +1)$. This factor
explicitly depends on the phonon dispersion law, but again, the major
contribution comes from short-wavelength high-frequency phonons, which
are determined by the short-range order in the electron system.

\section{Enhancement of tunneling by a magnetic field}

In this section we describe a new effect, the enhancement of the
tunneling rate by a magnetic field parallel to the electron
layer. Qualitatively, the enhancement is due to transferring the
energy of thermal in-plane motion into the energy of out-of-plane
tunneling. On the formal level it is a consequence of the increase,
with increasing temperature, of the absolute value of the term ${\cal
R}_{\rm ee}$ (\ref{retarded}) in the tunneling action. Since this term
gives a negative contribution to the tunneling exponent $R$, the whole
$B$-dependent term in ${\cal R}$ becomes negative starting with a
certain crossover temperature $T_c$, and then the tunneling rate
increases with $B$. The range boundaries where the overall escape rate
increases with $B$ are not universal and depend on the potential
$U(z)$ and the phonon spectrum. The enhancement occurs in a limited
temperature range, and may start from $B=0$ or have a finite threshold
in $B$. However, very strong fields suppress rather than enhance
escape.

\subsection{The crossover temperature}

The lower temperature bound of the enhancement domain is the
crossover temperature $T_c$. It can be determined from the small-$B$
expansion of the tunneling exponent for the ground state [$n=g$ in
Eq.~(\ref{general_W})],
\begin{equation}
\label{expansion}
R_g(\omega_c)\approx R_g(0)+A_g(T)\omega_c^2, \quad \omega_c\tau_0 \ll 1
\end{equation}
where $\tau_0$ is the tunneling time in the ground state for
$B=0$. The role of the ground state is special in that the barrier
width is bigger for the ground state energy than for the energies of
the excited states. Therefore the effect of the magnetic field, which
accumulates under the barrier, is most pronounced in the ground state.

The value of $A_g$ is given by the terms $\propto \omega_c^2$ in the
action $R_g$ (\ref{tun_Exp}) calculated along the tunneling trajectory
$z_0(\tau)$ for $B=0$.  From the analysis in Sec.~IV~A it
follows that $A_g>0$ for $T\to 0$. The crossover temperature is given
by

\begin{equation}
\label{T_c}
A_g(T_c)=0.
\end{equation}
For $T>T_c$ the tunneling exponent $R_g$
decreases and the tunneling rate increases with $B$, for small
$B$. 

In the limit of low phonon frequencies, $\omega_{{\bf k}j}\ll
1/\tau_0, T_c$, from Eqs.~(\ref{tun_Exp}), (\ref{classical}) it
follows that $\beta_c\equiv 1/T_c= 2\tau_0\overline {z_0}^2/\overline
{z_0^2}$, where $\overline {z_0}$ is the average coordinate $\bar z$
(\ref{classical}) for the $B=0$ trajectory with energy $E_g$, and
$\overline{z_0^2}$ is the mean square value of $z$ on the same
trajectory,
\[\overline {z_0^2} =
\tau_0^{-1}\int_0^{\tau_0} d\tau z_0^2(\tau)\quad (E=E_g).\] 
Clearly, in this case $\beta_c< 2\tau_0$. It follows from
Eq.~(\ref{retarded}) that $2\tau_0$ is also the limiting value of
$\beta_c$ in the opposite case of high phonon frequencies,
$\omega_{{\bf k}j}\gg 1/\tau_0$. On the whole, we have the bounds on
temperature for the tunneling enhancement in the ground intrawell
state
\begin{equation}
\label{claws}
2\tau_0 {\overline {z_0}^2 \over \overline {z_0^2}}< \beta_c <
2\tau_0.
\end{equation}
As noted above, $\overline{z_0}$ is nonzero, and generally $\overline
{z_0}^2 / \overline {z_0^2}\sim 1$. 

It follows from the above arguments that the value of the crossover
temperature $T_c =1/\beta_c$ decreases with increasing phonon
frequencies, that is the crossover is determined by high-frequency
phonons which, in the case of 2D electron systems, have large wave
numbers and are determined by the short-range order.

\subsection{Upper temperature limit}

A thresholdless tunneling enhancement starting from $B=0$, occurs for
temperatures bounded from above by the condition that the system
tunnels from the ground state rather than from excited
intrawell states or via thermal activation over the barrier. In
principle, even for excited states, the tunneling rate may increase
with $B$, but this does not happen for simple model potentials
investigated below.

If the tunneling is enhanced only in the ground state, the upper
temperature bound is often the temperature $T_{1\to 2}$ where the
probability of tunneling from the first excited state, weighted with
the occupation factor, exceeds that from the ground state, for $B=0$.
It can be estimated for smooth tunneling barriers, where the tunneling
duration $\tau_0(E)$ for $B=0$ often decreases with the increasing
energy $E$. In fact, the function $\tau_0(E)$ may be nonmonotonic even
for simple potentials $U(z)$; a detailed analysis of this function
lies outside the scope of this paper, but generalization of the
results to appropriate cases is straightforward. From (\ref{tun_Exp}),
for decreasing $\tau_0(E)$, switching from tunneling from the ground
state ($n=1$) to that from the first excited state ($n=2$) occurs for
the reciprocal temperature
\[ \beta_{1\to 2}=2 {\int_{E_1}^{E_2} \tau_0(E) dE \over E_2 -E_1} \quad 
(E_1\equiv E_g).\] 
This value lies between $2\tau_0(E_2)$ and $2\tau_0(E_1)$. Depending
on the tunneling potential, $\beta_{1\to 2}$ can be smaller or larger
than $\beta_c$ (\ref{claws}). If a magnetic field does not increase the
rate of tunneling from the state $n=2$,  thresholdless tunneling
enhancement occurs for $T_c < T_{1\to 2}$.

Alternatively, for $B=0$ the system may switch to activated escape
over the barrier with increasing temperature for $T=T_a< T_c$.  The
thresholdless tunneling enhancement by the magnetic field does not
occur in this case. However, in this as well as in the previous case
there may still occur a $B$-induced enhancement of the escape rate,
starting with some nonzero $B$. We note that in the above arguments,
it was assumed that thermalization inside the well is faster than
electron escapes.

\subsection{Field-induced  
switching between the levels and from activation to tunneling}

Even in the temperature range $T > T_{1\to 2}$ a sufficiently strong
magnetic field can increase the tunneling rate, provided $T >
T_c$. This happens if the tunneling exponent for the ground state
$R_g(\omega_c)\equiv R_{n=1}(\omega_c)$ exceeds that in the first
excited state $R_{n=2}(\omega_c)$ and its zero-field value
$R_{n=2}(0)$. In a certain temperature range where $T > T_{1\to 2}$,
the tunneling rate for $B=0$ is determined by tunneling from the
excited state $n=2$. This rate decreases with increasing $B$ (the
tunneling exponent $R_{n=2}$ increases with $B$). For some $B$ the
exponents $R_{n=2}(\omega_c)$ and $R_{n=1}(\omega_c)$ become equal to
each other. For larger $B$ the system tunnels from the ground state,
and the tunneling rate increases with $B$.

Similarly, since the activation rate is only weakly affected by $B$,
in a certain temperature range where escape already occurs via
activation for $B=0$, starting with some $B$ it may again go through
tunneling from the ground state. This happens if the tunneling rate
for the ground state becomes bigger than the activation rate and only
happens in a limited range of $B$, see Sec.~VI. For a special model
the switching is illustrated in Fig.~\ref{fig:switching} below.

\section{Tunneling enhancement for the Einstein model of a
Wigner crystal}

In what follows we will illustrate the general results and apply them
to specific 2D systems assuming that all
vibrational modes have the same frequency, $\omega_{{\bf k}j}=
\omega_p$, i.e. using the Einstein model of the Wigner crystal. This is
motivated by the fact that the tunneling is determined primarily by
short-wavelength vibrations, which have a comparatively weak
dispersion. Correspondingly, when we discuss the experiment, we will
set $\omega_p$ equal to the characteristic
short-wavelength plasma frequency $(2\pi e^2n^{3/2}/m)^{1/2}$, where
$n$ is the electron density.

\subsection{Triangular barrier}

For electrons above helium surface and in certain types of
semiconductor heterostructures, the potential $U(z)$ in the barrier
region ($z\geq 0$) is determined by the electric field which pulls
electrons away from the intrawell states. To a good approximation
$U(z)$ is then linear in $z$,
\begin{equation}
\label{triangular}
U(z)= {\gamma ^2 \over 2m }\left( 1- {z \over L}\right) \quad (z \geq 0).
\end{equation}
Here, $\gamma\equiv \gamma_1$ is the decrement of the ground-state
wave function $\psi_g\equiv \psi_1$ near the well, $\partial
\ln\psi_1/\partial z = -\gamma$ for $z=0$, cf. the discussion before
Eq.~(\ref{initial1}).  The additive constant in $U(z)$ is chosen so
that the energy of the ground state $E_g=0$. Then $L$ is the tunneling
length in the ground state for $B=0$. It is determined by the pulling
electric field. We assume that $\gamma L\gg 1$.

The approximation (\ref{triangular}) applies only within the barrier
region, where $U$ is determined by the pulling electric field, and not
inside the well, where $U(z)$ is singular. Moreover, it holds provided
the width of the tunneling barrier is small compared to the in-plane
interelectron distance $n^{-1/2}$ [cf. Eq.~(\ref{U_on_helium})
below]. 

In order to calculate the ground-state tunneling exponent, it is
convenient to solve directly the equations of motion (\ref{eom}) with
the boundary conditions (\ref{initial1}), (\ref{initial_p_u}),
(\ref{bound_Euclid}), and (\ref{tau}). For a triangular potential,
these equations are linear. This allowed us to obtain for the
tunneling exponent a simple expression
\begin{eqnarray}
\label{exact_Exp}
&\tilde{R}=&-\nu_p^2 \tau_{\rm red}^3 + 3\nu_p \tau_{\rm
red}(1-\tau_{\rm red})
\coth [\omega_p\beta/2-\nu_p\tau_{\rm red}] \nonumber \\
&&+3 +3\tau_{\rm red}(\nu^2-1), \quad R_g=2 \gamma L 
\tilde{R}/ 3 \nu^2.
\end{eqnarray}
Here, $\nu_p=\omega_p\tau_0$ and $\nu_c=\omega_c\tau_0$ are the
dimensionless in-plane and cyclotron frequencies scaled by the
tunneling duration $\tau_0$ for $B=0$, and $\nu^2=\nu_p^2 +\nu_c^2$.

The quantity $\tau_{\rm red}=\tau_f/\tau_0$ in Eq.~(\ref{exact_Exp}) is the
reduced tunneling duration. It is given by the equation
\begin{eqnarray}
\label{tauf}
&&\left[
(1-\tau_{\rm red})\nu_p\nu^2\coth [\omega_p\beta/2 -
\nu_p\tau_{\rm red}] -\nu_c^2\right] \tanh
\nu\tau_{\rm red} \nonumber \\
&&=\nu[\nu_p^2\tau_{\rm red} -\nu^2]
\end{eqnarray}

In the limit $T\to 0$, Eqs.~(\ref{exact_Exp}), (\ref{tauf}) go over
into the result obtained earlier \cite{us} (in Ref.~\cite{us} we used
$\omega_0$ and $\nu_0$ instead of $\omega_p$ and $\nu_p$,
respectively). In this limit, the role of the many-electron effects is
particularly important. In the single-electron approximation
($\omega_p=0$) the tunneling duration $\tau_f$ and the tunneling
exponent $R_g$ diverge for $\omega_c\to \tau_0^{-1}$
\cite{Andrei-93}. This happens because the effective single-electron
potential $U(z)+(1/2)m\omega_c^2z^2$, which takes into account the
parabolic magnetic barrier, does not have classically allowed extended
states with energy $E_g=0$ behind the barrier.

\begin{figure}
\begin{center}
\epsfxsize=3.0in                
\leavevmode\epsfbox{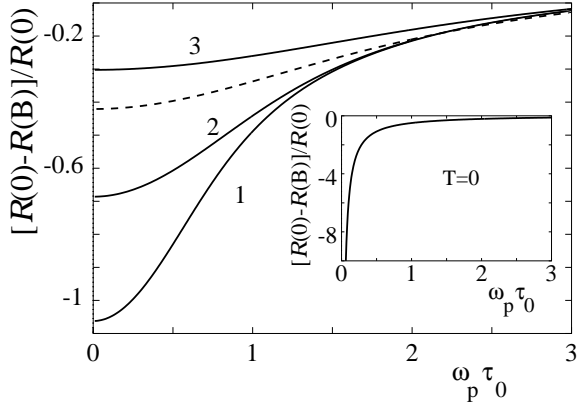}
\end{center}
\caption{The tunneling exponent in the ground state 
for a triangular potential barrier \protect(\ref{triangular}) as a
function of the phonon frequency $\omega_p$ in the Einstein model of
the Wigner crystal for $\omega_c\tau_0 = 2$. The time $\tau_0=
mL/\gamma$ is the duration of tunneling for $B=0$ and $T=0$. The
curves 1 to 3 refer to reciprocal temperatures $\beta/\tau_0 =7,\, 5,\,
3$. The dashed line is the result of the direct variational method, with
one variational parameter $\tau_f$.}
\label{fig:triangular_ground}
\end{figure}

The inter-electron momentum exchange makes tunneling possible for all
$B$. For $\omega_c\tau_0 > 1$ and $T=0$, the tunneling exponent is a
steep function of the exchange rate $\omega_p$ in the limit of slow
exchange, $\omega_p\tau_0\ll 1$.  In the opposite limit of the fast
momentum exchange, $\omega_p\gg
\tau_0^{-1} , \omega_c$, from Eqs.~(\ref{exact_Exp}), (\ref{tauf}) we
obtain $\tau_{\rm red}\approx 1$  [i.e., $\tau_f\approx
\tau_0$], and $R_g\approx 4\gamma L/3$. These are 
the values for tunneling for $B=0$. The overall dependence of the
tunneling exponent on $\omega_p$ for $T=0$ is shown in the inset of
Fig.~\ref{fig:triangular_ground}.

For a given magnetic field, the dependence of the tunneling exponent
$R_g$ on the frequency $\omega_p$ becomes much less steep with
increasing temperature, as seen from
Fig.~\ref{fig:triangular_ground}. For large $\omega_p\tau_0,\,
\omega_p\beta$, the curves for different temperatures merge together 
and approach the $B=0$ asymptote.

The value of $R_g$ can be calculated independently from the
functional ${\cal R}_n$ (\ref{tun_Exp}) using the direct variational
method. Even a simple approximation where $z(\tau)$ is quadratic in
$\tau$, with the only variational parameter being the tunneling
duration $\tau_f$, gives a reasonably good result, which is shown in
Fig.~\ref{fig:triangular_ground} by a dashed line for $\beta =
3\tau_0$. Such calculation gives a good approximation for higher
temperatures, and also for lower temperatures but not too small
$\omega_0\tau_0$. For low temperatures
and small $\omega_0\tau_0$ the trajectory $z(\tau)$ is strongly
nonparabolic, and more then one parameter is required in the
variational calculation.

\subsection{Field-induced tunneling enhancement and 
switching to tunneling from activation}

The explicit expression for the tunneling exponent (\ref{exact_Exp})
allows us to analyze the effects of tunneling enhancement and magnetic
field induced switching to tunneling, which were discussed in
Sec.~V. In the small-$B$ limit, where $\omega_c\ll \omega_p,
\tau_0^{-1}$, the tunneling exponent $R_g(B)$ is seen from
Eq.~(\ref{exact_Exp}) to be quadratic in $B$. The coefficient $A_g$ in
Eq.~(\ref{expansion}) can be easily calculated. From the condition
$A_g=0$ we obtain the value of the reciprocal temperature $\beta_c$
which corresponds to the crossover from decrease to increase of the
tunneling rate due to a magnetic field,
\begin{equation}
\label{betac}
\beta_c=2\tau_0+{2\over \omega_p}\tanh^{-1}
\left[{\nu_p[3\nu_p-(3+\nu_p^2)\tanh\nu_p] \over
\nu_p^3-3\nu_p + 3\tanh\nu_p}\right].
\end{equation}
In agreement with (\ref{claws}), $\beta_c$ monotonically increases
with $\omega_p$ from $5\tau_0/3$ at $\omega_p=0$ to $2\tau_0$ for
$\omega_p\to\infty$.

\begin{figure}
\begin{center}
\epsfxsize=3.35in                
\leavevmode\epsfbox{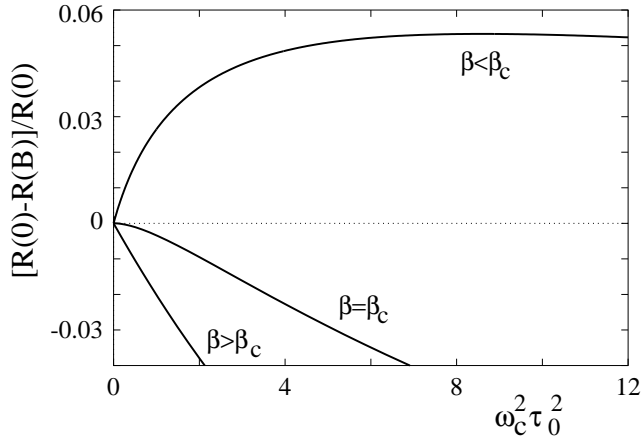}
\end{center}
\caption{The dependence of the tunneling exponent $R(B)\equiv R_g(B)$ on the 
magnetic field \protect(\ref{exact_Exp}) for $\omega_p\tau_0=1/3$ near
the crossover temperature $\beta_c \approx 1.67\tau_0$
\protect(\ref{betac}). The curves 1 to 3 correspond to 
$(\beta-\beta_c)/\tau_0 = 0.2, 0, -0.3$}
\label{fig:single-level}
\end{figure}

The dependence of the tunneling exponent (\ref{exact_Exp}) on the
magnetic field for different temperatures is shown in
Fig.~\ref{fig:single-level}. Above the crossover temperature $(\beta <
\beta_c$), $R(B)$ {\it decreases}, and $R(0)-R(B)$ and 
the tunneling probability {\it increase} with the increasing field,
for small $B$. The slope $dR/BdB \propto \beta - \beta_c$ for $B\to
0$.  However, for strong fields the tunneling rate  decreases
with the increasing $B$, because the Hall momentum can no longer be
compensated by thermal fluctuations. For small $\omega_p\tau_0$ this
happens when the typical Hall momentum $m\omega_cL$ becomes comparable
to the thermal momentum $(2mT)^{1/2}$ multiplied by the small factor
$(\gamma L)^{-1/2}$. This factor comes from the fact that the optimal
value of the transferred in-plane momentum in the single-electron
approximation is determined by the maximum of the sum $2S_E[p] +
(p^2/2mT)$, where $S_E[p] \propto \gamma L$ is the single-electron
action in the magnetic field for given in-plane momentum
$p$. Therefore the thermal momentum is scaled by $ (\gamma L)^{1/2}$.

It is clear from the data in Fig.~\ref{fig:single-level} that, for the
barrier chosen, the magnetic field induced increase of the tunneling
exponent $R$ is numerically small. However, for typical $R\agt 50$ it
can still be noticeable, although strictly speaking it is on the
border of applicability of the approximation in which only the
exponent is taken into account.

\begin{figure}
\begin{center}
\epsfxsize=3.35in                
\leavevmode\epsfbox{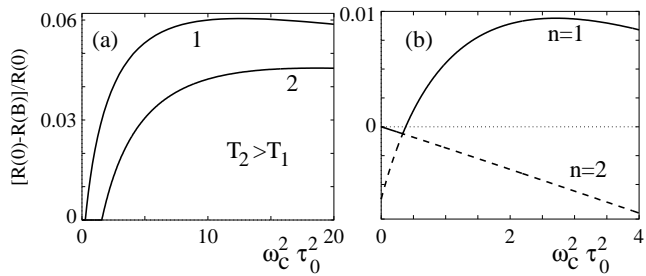}
\end{center}
\caption{Magnetic field induced switching from activation (a) and from tunneling from the excited state (b) to tunneling 
from the ground state, for $\omega_p\tau_0=1/3$. In (a), there is only
one intrawell state in the potential well $U(z)$, and the transition
to activation for $B=0$ occurs for $\beta/\tau_0=4/3$. The curves 1, 2
correspond to $(\beta-\beta_c)/\tau_0=-0.35, -0.4$. In (b), the
position $E_2$ of the excited level ($n=2$) is chosen at
$0.2\gamma^2/2m$ below the barrier top ($E_1=0$). The temperature is
chosen at $(\beta-\beta_c)/\tau_0=-0.16$, so that for $B=0$ the system
tunnels from the excited state. The observable (smaller) tunneling
exponents for a given $B$ are shown with bold lines, whereas dashed
lines show the bigger exponents, which correspond to smaller tunneling
rates.}
\label{fig:switching}
\end{figure}

The expression (\ref{exact_Exp}) gives the tunneling exponent only for
low enough temperatures where the system escapes from the ground
state. For higher temperatures, one should take into account the
possibility of escape from excited states and via an activated
transition over a potential barrier. The positions of the excited
levels depend not only on the barrier shape, but also on the shape of
the potential $U(z)$ inside the well. The analysis for a realistic
system, electrons on the surface of liquid helium, is done in the next
section. Here, in order to illustrate different options, we discuss
two cases: a narrow well, in which case the ground state is
essentially the only intrawell state, and a well with a comparatively
shallow excited state. We assume that the intrawell relaxation rate is
higher than the escape rate.

We start with the case of one bound state in the potential well. For
$B=0$ and a triangular barrier $U(z)$ (\ref{triangular}), switching
from tunneling to activation occurs here for the temperature
$T_a\equiv 1/\beta_a = (4\tau_0/3)^{-1}$. This temperature is higher
than the crossover temperature $1/\beta_c$ (\ref{betac}), and
therefore there is a region where the enhancement of tunneling by a
magnetic field can be observed, as discussed above
(cf. Fig.~\ref{fig:single-level}). However, even though for $T > T_a$
the $B=0$-escape occurs via overbarrier transitions, the increase of
the tunneling rate with the increasing $B$ can make tunneling more
probable for sufficiently strong $B$. If the activation rate is
independent of $B$, the overall dependence of the exponent of the
escape rate $R(B)$ on $B$ is shown in Fig.~\ref{fig:switching}a. In
this case, $R(0)=\gamma^2/2mT$ is the barrier height over
temperature. Switching to tunneling and to the increase of the escape
rate with $B$ occurs where the tunneling exponent $R_g(B)$ as given by
Eq.~(\ref{exact_Exp}) becomes less than $R(0)$.

A similar switching occurs in the temperature range where tunneling
from the first excited level is more probable than from the ground
state, for $B=0$. Since with increasing $B$ the tunneling rate in the
ground state increases, the system switches to tunneling from the
ground state starting with a certain value of $B$. This is illustrated
in Fig.~\ref{fig:switching}b.

In narrow-well potentials, a magnetic field may strongly affect the
wave functions with energies close to the barrier top. As a result,
new bound metastable states may appear in a strong field. The field
also shifts the energy levels of the existing states. The rate of
interlevel transitions may also change, since the field mixes together
the in-plane and out-of-plane motions. The related effects may become
important with increasing temperature.

\subsection{Square barrier: field-induced crossover to thermal 
activation}

In many physically interesting systems, the tunneling barrier $U(z)$
is nearly rectangular. This is often the case for semiconductor
heterostructures, where the barrier is formed by the insulating
layer. If we count $U$ off from the intrawell energy level $E_g$ and
set the boundaries at $z=0$ and $z=L$,
the barrier has the form
\begin{equation}
\label{rectan}
U(z)=\gamma^2/2m, \quad 0<z<L.
\end{equation}
Here, $1/\gamma$ is the decay length under the barrier,
cf. Eq.~(\ref{initial1}), and we have neglected the lowering of the
barrier due to the electrostatic field from other electrons at their
lattice sites, which is a good approximation for $nL^2\ll 1$.

We assume that, behind the barrier ($z> L$), an electron can move
semiclassically with all energies. Then the decaying underbarrier wave
function has to be matched to an appropriate propagating wave behind
the barrier at $z=L$.  In contrast to the case of a smooth barrier,
because the potential $U(z)$ is discontinuous at $z=L$, the
$z$-component of the momentum should not be the same on the opposite
sides of the boundary. However, the in-plane ``momentum'' components
${\bf u}_{{\bf k}j}$, which are imaginary under the barrier, still
have to be continuous. Respectively, the boundary conditions
(\ref{bound_Euclid}) for the tunneling trajectory should be changed to
\begin{equation}
\label{initial2}
z(\tau_f)=L, \quad {\bf u}_{{\bf k}j}(\tau_f)={\bf 0}.
\end{equation}
In fact, the condition ${\bf u}_{{\bf k}j}={\bf 0}$ gives the in-plane
values of ${\bf p}_{{\bf k}j}$ for which the wave function is maximal
for $z=L$.  

With the boundary conditions (\ref{initial2}), elimination of phonon
variables from the Euclidean action $S_E$ in the tunneling exponent is
similar to what was done for a smooth barrier. The resulting
expression for the retarded functional ${\cal R}_n[z]$ coincides with
Eq.~(\ref{tun_Exp}), provided $z(\tau_f+x)$ is defined as $z(\tau_f
-x)$, for $0\leq x\leq \tau_f$.

An important feature of a rectangular tunneling barrier is that the
tunneling time $\tau_0(E)=-dS_0/dE$ for $B=0$ monotonically increases
with energy $E$. Therefore the maximum of the function $-\beta E -
2S_0(E)$, which gives the probability of tunneling with energy $E$
with account taken of the occupation factor, corresponds either to the
transition from the ground state or to activation over the
barrier. Switching to activation occurs for the temperature
$T_a=\gamma^2/4mS_0(E_g)\equiv \gamma/4mL= (4\tau_0)^{-1}$. It is
lower than the temperature $T_c$ of the crossover from $B$-suppressed
to $B$-enhanced tunneling as given by Eq.~(\ref{claws}), and therefore
we do not expect the crossover to occur in systems with a
square barrier.

If the temperature $T< T_a$, escape for $B=0$ occurs via tunneling,
and its probability decreases with the increasing $B$. Starting with
some $B$, where the tunneling exponent becomes bigger than the
activation exponent $\gamma^2/2mT$, it becomes more probable to escape
by activated transition than by tunneling. To a good approximation,
the escape rate becomes than independent of the magnetic field.

\begin{figure}
\begin{center}
\epsfxsize=3.35in                
\leavevmode\epsfbox{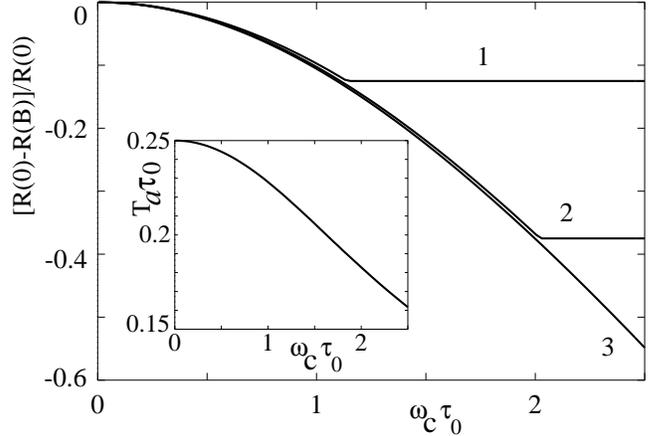}
\end{center}
\caption{The logarithm of the escape rate $R(B)$ for the square 
potential barrier, $\omega_p=(2\tau_0)^{-1}$ ($\tau_0=
mL/\gamma$). The value $R(0)$ is given by the tunneling exponent,
$R(0)=2S_0(E_g)\equiv 2\gamma L$. Curves 1 to 3 correspond to
$\beta/\tau_0= 4.5,5.5,6.5$. The sections of the curves where $R(B)$
increases correspond to tunneling and are described by
Eq.~(\protect\ref{tun_square_pot}), whereas the horizontal sections of
the curves correspond to thermal activation. Inset: the magnetic field
dependence of the switching temperature $T_a\tau_0$.}
\label{fig:semi.eps}
\end{figure}

The $B$-dependence of the escape rate for different temperatures is
illustrated in Fig.~\ref{fig:semi.eps}. The results refer to the
Einstein model of the Wigner crystal. In this model the tunneling
exponent can be obtained directly from the [linear, in this case]
equations of motion (\ref{eom}) with the boundary conditions
(\ref{initial1}), (\ref{initial_p_u}), (\ref{initial2}). It has the
form

\begin{equation}
\label{tun_square_pot}
R_g=\gamma L\left[1+\tau_{\rm red}+\nu_c \kappa(\tau_{\rm red})\right],
\end{equation}
where the function $\kappa(\tau_{\rm red})$ and the reduced tunneling
time $\tau_{\rm red}=\tau_f/\tau_0$ are given by the equations

\wid
\begin{eqnarray}
\label{trd_square_pot}
\kappa (\tau_{\rm red}) \equiv && {\nu_c(\cosh\nu\tau_{\rm red}-1)
\over \nu_c^2+\nu_p^2 \cosh \nu\tau_{\rm red} + \nu \nu_p 
\coth[\omega_p\beta/2-\nu_p\tau_{\rm red}] 
\sinh \nu \tau_{\rm red}}\nonumber\\
&& ={1\over\nu_c\nu_p^2} {\nu_c^2(2-2\cosh\nu\tau_{\rm rd}+\nu\tau_{\rm rd}
 \sinh\nu\tau_{\rm rd})-\nu^3(\tau_{\rm rd}-1)\sinh\nu\tau_{\rm rd}
 \over (1-\cosh\nu\tau_{\rm rd})(1-\nu_c^2/\nu_p^2)+\nu\tau_{\rm rd}
 \sinh \nu\tau_{\rm rd}}
\end{eqnarray}
\nar

\noindent
with $\nu_p=\omega_p\tau_0, \nu_c=\omega_c\tau_0$, and
$\nu=(\nu_p^2+\nu_c^2)^{1/2}$. 

The temperature of switching to activation is given by the equation
$T_a= \gamma^2/2mR_g$. From (\ref{tun_square_pot}),
(\ref{trd_square_pot}), both $R_g$ and the reduced time $\tau_{\rm
red}$ increase with the magnetic field, and therefore the switching
temperature $T_a$ decreases with $B$. However, it follows from the
analysis of the above equations that $T_a$ remains lower than
$1/(2\tau_f)$, and $\tau_f \geq \tau_0$ for all temperatures where the
system escapes via tunneling, in contrast to the case of the
triangular barrier discussed earlier.

The effect of saturation of the escape rate with increasing $B$ is not
limited to square barriers, of course. For strong enough $B$ and
nonzero temperatures, the tunneling rate becomes less than the
activation rate, and the system switches to activation; the switching
may go in steps with increasing $B$, via tunneling from excited
intrawell states.

\section{Comparison with the experimental data on tunneling 
from helium surface }

Tunneling from a strongly correlated 2DES has been investigated in
much detail for electrons on helium surface \cite{Goodkind,Andrei-93}.
In this system, a good agreement has been reached between theory and
experiment in the absence of the magnetic field, where the primary
role of the electron correlations is to change the effective
single-electron tunneling barrier (see below). As mentioned before,
there were also done interesting experiments on tunneling in a
magnetic field. However, the observed strong field dependence of the
tunneling rate dramatically differed from the predictions of the
single-electron theory and has remained unexplained
\cite{Andrei-93}.

Electrons on helium surface are localized in a 1D potential box. One
side of this box is the image potential $-\Lambda/z$, where $\Lambda =
e^2(\epsilon-1) /4(\epsilon+1)$ ($\epsilon\approx 1.057$ is the
dielectric constant of helium), and $z$ is the direction normal to the
surface. The other side is a high barrier $\sim 1$~eV on the surface
($z=0$), which prevents electrons from penetrating into the
helium. The intrawell states can be made metastable by applying a
field ${\cal E}_{\perp}$ which pulls the electrons
away from the surface.  This field is determined by the helium cell
geometry and depends on the applied voltage and the electron density
$n$, cf.~\cite{Eperp}. The overall electron potential has the form

\begin{equation}
\label{U_on_helium}
U(z) = -\Lambda z^{-1} - |e{\cal E}_{\perp}|z -m\bar\omega^2z^2 \quad (z>0).
\end{equation}

The term $\propto \bar\omega^2$ describes the Coulomb
field created by other electrons at their in-plane lattice sites
(the ``correlation hole''
\cite{AP-90,Iye-80}). Only the lowest-order term in the ratio of the
tunneling length $L$ to the interelectron distance $n^{-1/2}$ has been
kept in Eq.~(\ref{U_on_helium}), and we used the interrelation

\begin{equation}
\label{omega_0}
\bar\omega=\left[{e^2\over 2m}{\sum_l}^{\prime}|{\bf R}_l|^{-3}\right]^{1/2} 
\equiv \left[{1\over 2N}\sum_{{\bf k}j}
\omega^2_{{\bf k}j}\right]^{1/2}
\end{equation}
for the sum over
lattice sites ${\bf R}_l$. For a triangular lattice, $\bar\omega
\approx \left(4.45 e^2n^{3/2}/ m\right)^{1/2}$ \cite{Mark}. 
The conditions $1/\gamma\ll L\ll n^{-1/2}$
are typically very well satisfied in experiment, with the decay length
$1/\gamma = 1/\Lambda m \approx 0.7\times 10^{-6}$~cm, $L\sim
|E_g/e{\cal E}_{\perp}|\approx \gamma^2/2m|e{\cal E}_{\perp}|\sim
10^{-5}$~cm, and $n^{-1/2} \sim 10^{-4}$~cm [in the estimate of $L$ we
used that $|E_g|\gg |e{\cal E}_{\perp}|/\gamma,
m\bar\omega^2/\gamma^2$, and that $|e{\cal E}_{\perp}|/\gamma \agt
\bar\omega$].

To compare the predicted dynamical effect of the electron-electron
interaction with the experimental data on tunneling in the magnetic
field \cite{Andrei-93}, we use the Einstein model of the WC. In the
equations of motion (\ref{eom}) we assume that all phonon frequencies
$\omega_{{\bf k}j}$ are the same and set them equal to the
characteristic plasma frequency $\omega_p=(2\pi e^2n^{3/2}/m)^{1/2}$
[yet we use Eq.~(\ref{omega_0}) for the parameter $\bar\omega$ of the
potential barrier $U(z)$]. The numerical results change only slightly
when the phonon frequency is varied within reasonable limits, e.g., is
replaced by $\bar\omega$.

\begin{figure}
\begin{center}
\epsfxsize=3.35in                
\leavevmode\epsfbox{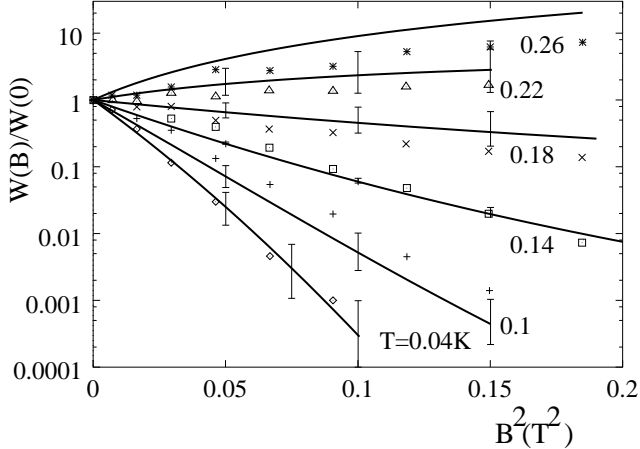}
\end{center}
\caption{The relative rate of electron tunneling  from the 
helium surface $W(B)/W(0)$ as a function of the magnetic field $B$ for
the electron density $n=0.8\times 10^8 cm^{-2}$ and the calculated
pulling field ${\cal E}_{\perp}=24.7$~V/cm (solid curve). Solid lines
show how the theory compares to the experimental data points of
Ref.~\protect\cite{Andrei-93}. The error bars show the uncertainty in
the theoretical values due to the uncertainty in the parameters of the
experiment.}
\label{fig:exponent}
\end{figure}

The magnetic field dependence of the tunneling rate for the parameters
used in the experiment is calculated from Eqs.~(\ref{eom}),
(\ref{initial1}), (\ref{bound_Euclid}), (\ref{initial_p_u}) and is
shown in Fig.~\ref{fig:exponent}. The data refer to the values of $T$
where escape occurs via tunneling from the ground state. The actual
calculation is largely simplified by the fact that, deep under the
barrier, the image potential $-\Lambda/z$ in (\ref{U_on_helium}) can
be neglected. The equations of motion (\ref{eom}) become then linear,
and the tunneling exponent $R$ can be obtained in an explicit,
although cumbersome form, which was used in
Fig.~\ref{fig:exponent}. The correction to $R$ from the image
potential is $\sim 1/\gamma L$, which is the small parameter of the
theory. Moreover, since this correction comes from the range of small
$z$, where the effect of the magnetic field is small, it is largely
compensated where $R(B)-R(0)$ is calculated.  This and other
corrections $\sim 1/\gamma L$ result in changes of the theoretical
curves that are smaller than the uncertainty in $R(B)-R(0)$ due to the
uncertainties in $n$ and ${\cal E}_{\perp}$ in the experiment
\cite{Andrei-93}.

As seen from Fig.~\ref{fig:exponent}, the dynamical many-electron
theory is in good qualitative and quantitative agreement with the
experiment, without any adjustable parameters. At low temperatures
($T=0.04$~K), the many-electron tunneling rate is bigger than the
single-electron estimate \cite{Andrei-93} by a factor of 10$^2$ for
$B=0.25$~T. For this temperature, the tunneling rate is well described
by the $T\to 0$ limit \cite{yet_to_be}. The $B$-dependence of the
tunneling rate is very sensitive to temperature. It becomes less
pronounced for higher $T$, and the role of dynamical many-electron
effects becomes less important, too. Interestingly, the theoretical
data on the {\it ratio} of $W(B)/W(0)$ become less sensitive to the
experimental uncertainties in the cell geometry (which determines
${\cal E}_{\perp}$) and the electron density $n$ for intermediate
temperatures $T\sim 0.14$~K. This is because the corresponding errors
in $W(B)$ and $W(0)$ compensate each other for such temperatures.

The crossover to magnetic-field enhanced tunneling occurs for
temperature $T_c\approx 0.19 K$, for the parameters in
Fig.~\ref{fig:exponent}. The expected increase of the tunneling rate
with $B$ for $T> T_c$ is shown in  Fig.~\ref{fig:exponent}. It has
indeed been observed in the experiment \cite{Andrei-93}. The
analysis of the experiment requires to establish whether, for
temperatures of interest, escape actually occurs via tunneling.  To
that end we note first that, as it follows from a direct variational
calculation, the potential $U(z)$ (\ref{U_on_helium}), with the
parameter values specified in Fig.~\ref{fig:exponent}, has only one
metastable intrawell state.

If the intrawell relaxation were fast enough, the temperature of the
crossover from tunneling to activation $T_a$ for $B=0$ would be given
by the condition that the tunneling exponent $R_g$ be equal to the
activation exponent $(U_{\max}-E_g)/T$ [here, $U_{\max}$ is the
maximal value of the potential $U(z)$]. This would give $T_a\approx
0.15$~K. However, activated escape requires that the in-plane thermal
energy of an electron be transformed into the energy of its
out-of-plane motion. This involves a large transfer of the
in-plane momentum $\sim [2m(U_{\max}-E_g)]^{1/2}$. The
electron-electron interaction does not give rise to such a transfer in
a strongly correlated system, since the reciprocal interelectron
distance is $n^{1/2} \ll [2m(U_{\max}-E_g)]^{1/2}$.

The major process which gives rise to the momentum transfer is
scattering by capillary waves on the helium surface, ripplons
\cite{Andrei_book}. Electron-ripplon coupling is weak. As a result, the 
prefactor in the activation rate, which is quadratic in the coupling
constant, is small. For $B=0$ it is $\sim\gamma^2T^2/\hbar\sigma$
\cite{Nagano-79},  where $\sigma$ is the surface tension of liquid helium.
For temperatures $T< 0.25$~K this prefactor is less than the prefactor
in the tunneling rate $(\hbar\gamma^2/m)\exp(-2)$ by a factor $<
10^{-5}$. Therefore the crossover from tunneling to activation
occurs for higher temperatures than it would follow from the condition
of equal tunneling and activation exponents.

For the parameters in Fig.~\ref{fig:exponent}, the rates of activation
and tunneling escape become equal for temperatures slightly higher
than 0.26~K (for $B=0$). Therefore the experimentally observed
increase of the escape rate with $B$ is indeed due to the discussed
mechanism of $B$-enhanced tunneling. The smaller experimental values
of the relative escape rate $W(B)/W(0)$ for $T=0.26$~K can be
understood by noticing that the activation rate is close to the
tunneling rate for such $T$, and since it presumably only weakly
depends on $B$\cite{B-activation}, the overall slope of
$\ln[W(B)/W(0)]$ should be smaller than that of the theoretical curve
which ignores activation (approximately, by a factor of 2, if we
ignore the dependence of the activation rate on $B$).

\subsection{The prefactor}

The dependence of the potential $U(z)$ (\ref{U_on_helium}) on $n$
gives rise to the density dependence of the tunneling rate $W(B)$ even
for $B=0$. We calculated the exponent and the prefactor in $W(0)$ by
matching the WKB wave function under the barrier for $1/\gamma \ll
z\ll L$ with the tail of the non-WKB intrawell solution (here,
$L=\hbar^2\gamma^2/2m|e{\cal E}_{\perp}|$ is the characteristic barrier
width). In the spirit of the logarithmic perturbation theory (LPT)
\cite{LPT}, the wave function of the ground state 
inside the well and not too far from it can be sought in the form
\begin{equation}
\label{lpt}
\psi_g(z)={\rm const}\times z\exp[-A(z)] 
\end{equation}
[we explicitly take into account that the
function $\psi_g(z)$ has a zero in the ground state].

The function $dA/dz$ satisfies a Riccati equation. It can be solved
near the well ($z\ll L$) by considering the last two terms in the
potential $U(z)$ (\ref{U_on_helium}) as a perturbation. For small $z$,
the major correction comes from the term $\propto {\cal
E}_{\perp}$. To the first order in ${\cal E}_{\perp}$,

\begin{equation}
\label{perturb_tail}
A(z) \approx \gamma z\left(1 - {z\over 4L}\right).
\end{equation}
In obtaining this expression we took into account the correction to
the ground state energy $\delta E_g=-3|e{\cal E}_{\perp}|/2\gamma$.
This correction can be obtained from the condition that the linear in
${\cal E}_{\perp}$ term in $dA/dz$ remain finite for $z\to 0$. 

The correction to $A$ (\ref{perturb_tail}) is small for $z$ small
compared to the barrier width $L$. We note that the exponent $A(z)$
has an overall functional form which differs from that of the commonly
used \cite{Andrei_book} variational wave function $\psi(z) \propto
z\exp(-\tilde\gamma z)$, with $\tilde\gamma$ being the variational
parameter. 

\begin{figure}
\begin{center}
\epsfxsize=3.0in                
\leavevmode\epsfbox{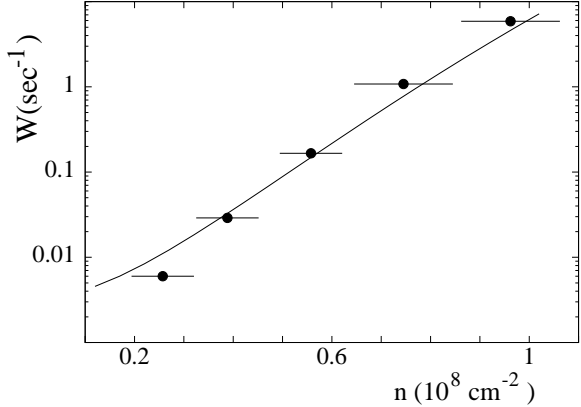}
\end{center}
\caption{The rate of electron tunneling  from the 
helium surface $W(0)$ for $B=0$ as a function of the electron density.
The dots show the experimental data \protect\cite{Andrei-93}. The
pulling field ${\cal E}_{\perp}$ was calculated from the geometry of
the experimental cell, the applied voltage, and the density.}
\label{fig:density_dependence}
\end{figure}

The expression for $A$ (\ref{perturb_tail}) matches the small-$z/L$
expansion of the action $S$ of the WKB wave function under the barrier
for $L\gg z\gg \gamma^{-1}$. This allowed us to find the prefactor in
the WKB wave function and in the tunneling rate.  The resulting
tunneling rate is shown in Fig.~\ref{fig:density_dependence}. It fully
agrees with the experiment (see also Ref.~\cite{Goodkind}).

\section{Conclusions}
It follows from the results of the present paper that interelectron
momentum exchange in correlated 2D electron systems leads to an
exponentially strong change of the rate of tunneling decay in a
magnetic field parallel to the electron layer. The mechanism is
dynamical by nature. It depends on the interrelation between the
characteristic momentum exchange rate and the reciprocal duration of
tunneling in imaginary time $1/\tau_f$. 

For low temperatures, where escape occurs via tunneling from the
ground state, the tunneling rate is affected primarily by
high-frequency in-plane electron vibrations. In their turn, these
vibrations are determined by short-range order in the 2DES. Their
frequencies are of the order of the characteristic zone-boundary
frequency of the Wigner crystal $\omega_p$. If $\omega_p\gg 1/\tau_f$,
the effect of the magnetic field on tunneling is nearly completely
compensated in the case where the width of the tunneling barrier is
small compared to the interelectron distance.

At higher temperatures, the magnetic field may in fact {\it enhance}
rather than suppress the rate of tunneling decay. The overall escape
rate as a function of $B$ and $T$ is expected to display a number of
other unusual features. These include switching from activated escape
to tunneling and vice versa, and switching between tunneling from the
ground and excited states. These switchings have been analyzed for
simple but realistic models of the tunneling barrier.

Our results on the field dependence of the tunneling rate and its
evolution with temperature are in full qualitative and quantitative
agreement with the existing experimental data on tunneling from a
strongly correlated 2DES on helium \cite{Andrei-93}, with no
adjustable parameters.

The results also apply to 2DES in semiconductor heterostructures. For
correlated systems in semiconductors, tunneling has been investigated
mostly for the magnetic field $\bf B$ perpendicular or nearly
perpendicular to the electron layer, cf. \cite{Spielman-00}. The data
on tunneling in a field parallel to the layer refer to high density
2DESs \cite{Eisenstein}, where correlation effects are small.

The effect of a parallel magnetic field is most pronounced in systems
with shallow and broad barriers $U(z)$. For example, in a GaAlAs
structure with a square barrier of width $L=0.1\, \mu$m and height
$\gamma^2/2m=0.02$~eV, for the electron density $n= 1.5\times 10^{10}$
cm$^{-2}$ and $B= 1.2$~T we have $\omega_p\tau_0
\approx 0.6$ and $\omega_c\tau_0 \approx 1$ ($\tau_0=mL/\gamma$ is the
tunneling duration for $n=B= 0$). The results of Sec.~VI C for square
barriers (with account taken of the correlation-hole correction) show
that the interelectron momentum exchange should significantly modify
the tunneling rate in this parameter range, provided the 2DES is
correlated \cite{yet_to_be}.  We therefore expect that tunneling
experiments on low-density 2DESs in parallel fields will reveal
electron correlations not imposed by the magnetic field, give insight
into electron dynamics, and possibly even reveal a transition from an
electron fluid to a pinned Wigner crystal with decreasing $n$.

We are grateful to B.I. Shklovskii for a helpful discussion. This
research was supported in part by the NSF through Grant
No. PHY-0071059.

\end{multicols}

\begin{thebibliography}{99}
\bibitem{B-activation} The rate of activation escape may depend 
on the magnetic field, because the field ``pushes'' the ground state
upward in energy, by $m\omega_c^2[\langle z^2\rangle -\langle
z\rangle^2]/2$, for a weak field (the averaging is performed for the
ground state).  Another factor is the change, by the magnetic field,
of the wave functions with energies close to the barrier top. For
$B=0.4$~T the magnetic length $l=(\hbar c/eB)^{1/2}$ is $\sim 0.6$ of
the distance from the helium surface to the barrier top position
$(\Lambda/|e{\cal E}_{\perp}|)^{1/2}$. Both factors lead to the decrease of
the activation energy of ecsape.
\end{thebibliography}
\end{document}